\documentclass{emulateapj}
\usepackage{graphicx}
\begin{document}

\def\etal{et al.\ \rm}
\def\ba{\begin{eqnarray}}
\def\ea{\end{eqnarray}}
\def\etal{et al.\ \rm}

\title{Properties of gravitoturbulent accretion disks}

\author{Roman R. Rafikov\altaffilmark{1,2}}
\altaffiltext{1}{Department of Astrophysical Sciences, 
Princeton University, Ivy Lane, Princeton, NJ 08540; 
rrr@astro.princeton.edu}
\altaffiltext{2}{Sloan Fellow}


\begin{abstract}
We explore the properties of cold gravitoturbulent accretion 
disks --- non-fragmenting disks hovering on the verge of gravitational 
instability --- 
using a realistic prescription for the effective viscosity 
caused by gravitational torques. This prescription is based on
a direct relationship between the angular momentum transport 
in a thin accretion disk and the disk cooling in a steady state. 
Assuming that opacity
is dominated by dust we are able to self-consistently derive disk
properties for a given $\dot M$ assuming marginal gravitational 
stability. We also allow external irradiation of the 
disk and account for a non-zero background viscosity which can 
be due to the MRI. Spatial transitions between different co-existing 
disk states (e.g. between irradiated and  self-luminous or between 
gravitoturbulent 
and viscous) are described and the location of the boundary at 
which disk must fragment is determined in a variety of situations. 
We demonstrate in 
particular that at low enough $\dot M$ external irradiation 
stabilizes gravitoturbulent disk against fragmentation all the way 
to infinity
thus providing means of steady mass transport to the central 
object.  Implications of our results for the possibility 
of planet formation by gravitational instability in protoplanetary 
disks and star formation in the Galactic Center and for the problem 
of feeding supermassive black holes in galactic nuclei are discussed.
\end{abstract}


\section{Introduction.}  
\label{sect:intro}

Gravitational instability (GI) in astrophysical disks has
been a subject of investigation for more than fifty years 
since the seminal work by Safronov (1960) and Toomre (1964). Originally 
it was studied predominantly in the context of driving the 
spiral structure in galaxies. Later it has been suggested
that GI may play important role in planet formation (Cameron 1978; 
Boss 1998), and
its significance for the properties of compact nuclear disks
around supermassive black holes in the centers of galaxies 
has also been recognized (Paczynski 1978a; Goodman 2003). 

It is generally accepted that the GI sets in when the so called
Toomre $Q$ satisfies the following condition:
\ba
Q\equiv\frac{\Omega c_s}{\pi G \Sigma}< Q_0,
\label{eq:Q}
\ea
where $\Sigma$, $\Omega$, and $c_s$ are the local 
surface density, angular frequency, and the sound speed in the
disk which we consider to be made of gas and having a Keplerian
rotational profile. $Q_0$ is a constant of order unity, its precise
value determining the instability threshold ranges from $0.7$ to
$1.7$ according to different authors (Kim \etal 2002; Boss 2002). 
Nonlinear development of the GI sensitively depends on the 
thermodynamical properties of the gas as has been 
first shown by Gammie (2001): if gas 
can cool on a timescale shorter than the local dynamical timescale 
$\Omega^{-1}$ then disk fragments into bound, self-gravitating 
objects which may grow further by accreting the surrounding gas and
colliding with each other. But if the cooling timescale $t_{cool}$
is longer than $\Omega^{-1}$ then the disk settles into a state
of the so-called {\it gravitoturbulence} in which surface density 
can experience significant fluctuations but the disk is stable 
against fragmentation in the long run and maintains itself on the brink
of instability with $Q\approx Q_0$. Stability against fragmentation 
arises because the restoring action of the thermal pressure 
resisting the self-gravity of overdensities is not sufficiently 
reduced by cooling when $t_{cool}\gtrsim \Omega^{-1}$.
 
Torques produced by the nonaxisymmetric density perturbations 
in gravitoturbulent 
disks give rise to angular momentum transport. Considering the
disk to be in a steady state on time shorter than its viscous 
evolution timescale one can directly relate ``effective viscosity''
$\alpha_{GI}$ 
produced by the gravitational torques to the cooling time of the disk. 
Gammie (2001) has demonstrated that the dimensionless $\alpha$-parameter
characterizing disk viscosity (Shakura \& Sunyaev 1973) is
\ba
\alpha\sim(\Omega t_{cool})^{-1}.
\label{eq:alpha}
\ea
It is instructive to show where this relation comes from. First, in a 
steady state the rate of energy dissipation per unit surface area 
of the disk $\sim \Omega^2\dot M$ has to equal the energy flux $F$
emitted from the disk surface. Second, the accretion rate is 
$\dot M=3\pi\nu\Sigma$, where $\nu\equiv\alpha c_s^2/\Omega$
(Pringle 1981). Combining these 
relations one immediately obtains equation (\ref{eq:alpha}) with
\ba
t_{cool}\approx \Sigma c_s^2/F.
\label{eq:t_cool_simple}
\ea
Relation (\ref{eq:alpha}) makes it possible to interpret disk 
fragmentation occurring at $t_{cool}\sim 
\Omega^{-1}$ as the inability of the disk to sustain gravitational
stress at $\alpha\gtrsim 1$ (Rice \etal 2005). 

In this paper we investigate the structure and evolution of
gravitoturbulent disks in which angular momentum is transferred 
predominantly by the gravitational torques. This problem has been 
previously investigated by Lin \& Pringle (1987) but with a rather 
naive prescription for the effective viscosity. Also, some 
efforts have been devoted to understanding the structure of the 
gravitoturbulent disks which are unstable to fragmentation on large 
scale, i.e. disks having $Q\approx Q_0$ and $\alpha_{GI}\sim 1$ 
everywhere (Rafikov 2005, 2007; Matzner \& Levin 2005). In this work 
viscous evolution of the disk is explored according to the prescription 
(\ref{eq:alpha}) without fixing the value of $\alpha_{GI}$ --- instead it
is calculated self-consistently based on the physical properties of the 
gas. We concentrate our attention on rather cool disks in which opacity
is due to dust grains thus focussing on the GI in the 
outer parts of protostellar disks and disks around supermassive black 
holes.

\section{General considerations.}
\label{sect:setup}

We consider a gravitoturbulent disk in which the dissipation
of transient density waves excited by GI is capable of maintaining 
$Q=Q_0$, and the cooling time $t_{cool}$ is longer than $\Omega^{-1}$. 
Cooling time of the disk is 
\ba
t_{cool}\approx
\frac{\Sigma c_s^2}{\sigma T^4}f(\tau),
\label{eq:t_cool}
\ea
where $\Sigma$ is the surface density of the disk, 
$c_s\equiv (k_B T/\mu)$ is the isothermal sound speed 
determined by the midplane temperature $T$, and $f(\tau)$ is
a function of the optical depth $\tau=\int \kappa\rho dz$ ($\kappa$
and $\rho$ are the gas opacity and density, $z$ is the vertical 
coordinate) which links the emitted flux $F$ to $T$: $F=\sigma T^4/f(\tau)$. 
A specific form 
of $f(\tau)$ depends on the way in which energy is transported from 
the midplane of an optically thick disk to its photosphere where 
it is radiated to space. 
Rafikov (2007) has calculated $f(\tau)$ in the case of efficiently 
convecting disks. However, in this work we assume (as was previously 
done in Rafikov 2005) for simplicity that  
energy is carried from the disk midplane to its surface solely by
radiation in which case $f(\tau)$ can be reasonably well approximated
by
\ba
f(\tau)\approx \tau+\frac{1}{\tau}.
\label{eq:f_tau}
\ea
This expression smoothly interpolates between the cooling rates 
applicable in the optically thick ($\tau\gg 1$) and  
optically thin ($\tau\ll 1$) regimes.

We assume a temperature-dependent opacity in the form
\ba
\kappa=\kappa_0 T^\beta,
\label{eq:opacity}
\ea
which is appropriate at low temperatures when $\kappa$ is dominated 
by dust grains. At very low temperatures, $T<150$ K, it is generally 
found (Bell \& Lin 1994; Semenov \etal 2003) that opacity is 
due to the icy grains and is characterized by 
\ba
\beta=2~~~ \mbox{and}~~~ \kappa_0\approx 5\times 10^{-4} \mbox{cm}^2 
\mbox{g}^{-1} \mbox{K}^{-2}
\label{eq:dust_opacity}
\ea
within 
a factor of 2 or so. At higher temperatures ices evaporate and
opacity behavior can be crudely described as $\kappa\approx 0.1T^{1/2}$
cm$^2$ g$^{-1}$ (Bell \& Lin 1994). 
For simplicity in this work we do not distinguish 
between the Rosseland mean and the Planck mean opacities (appropriate 
for $\tau\gg 1$ and $\tau\ll 1$ correspondingly) as they have 
similar values at low $T$.

Now, using definition (\ref{eq:Q}) and condition $Q=Q_0$ we find
that 
\ba
&& c_s=\frac{\pi G Q_0\Sigma}{\Omega},
\label{eq:c_s}\\
&& T=\frac{\mu}{k_B}\left(\frac{\pi G Q_0\Sigma}{\Omega}\right)^2
\label{eq:T}
\ea
in a gravitoturbulent disk. In the optically thick regime total
optical depth is dominated by the midplane layers of the disk in
which most of the mass is concentrated, so that up to 
factors of order unity $\tau\approx \kappa(T)\Sigma$. Clearly, this
approximation also works in the optically thin case. Thus,
using equation (\ref{eq:T}) one rather generally finds that
\ba
\tau\approx \kappa_0\Sigma^{2\beta+1}\left(\frac{\mu}{k_B}\right)^\beta
\left(\frac{\pi G Q_0}{\Omega}\right)^{2\beta}.
\label{eq:tau}
\ea

We can also calculate $\alpha_{GI}$ characterizing angular 
momentum transport caused by the non-axisymmetric surface 
density perturbations. Using equations
(\ref{eq:alpha}), (\ref{eq:t_cool}), and (\ref{eq:T}) one finds
that
\ba
\alpha_{GI}=\zeta \frac{\sigma (\pi G Q_0)^6}{f(\tau)}
\left(\frac{\mu}{k_B}\right)^4\frac{\Sigma^5}{\Omega^7}.
\label{eq:alpha_GI}
\ea
The kinematic viscosity $\nu\equiv \alpha_{GI}c_s^2/\Omega$ is then
given by the following expression:
\ba
\nu_{GI}=\zeta \frac{\sigma (\pi G Q_0)^8}{f(\tau)}
\left(\frac{\mu}{k_B}\right)^4\frac{\Sigma^7}{\Omega^{10}}.
\label{eq:nu}
\ea
Parameter $\zeta\sim 1$ appearing in these equations absorbs our
ignorance of the exact values of constant factors in
equations (\ref{eq:alpha}) and (\ref{eq:t_cool}).

Equations (\ref{eq:tau}), (\ref{eq:alpha_GI}) and (\ref{eq:nu}) 
provide us with the desired viscosity prescription needed 
for determining the physical structure and evolution of the 
gravitoturbulent disk. The only two essential ingredients that went
into deriving this viscosity recipe are (1) requirement that  
disk maintains itself in a state of marginal stability with respect 
to GI and (2) prescription (\ref{eq:alpha}) for $\alpha_{GI}$
which arises from reasonable assumption that disk is in thermal 
equilibrium on timescales shorter than the viscous timescale.

\section{Constant $\dot M$ disks.}
\label{sect:mdot}

In this section we consider the structure of gravitoturbulent 
disk with a specified mass accretion rate $\dot M$. In a steady state 
$\dot M = 3\pi \nu\Sigma$,
which with the aid of equation (\ref{eq:nu}) can be manipulated
into the following general relation:
\ba
\dot M = 3\pi\zeta \frac{\sigma (\pi G Q_0)^8}{f(\tau)}
\left(\frac{\mu}{k_B}\right)^4\frac{\Sigma^8}{\Omega^{10}}
\label{eq:rel1} 
\ea
Function $f(\tau)$ entering this expression depends on
$\Sigma$ and $\Omega$ through equation (\ref{eq:tau}). This
allows us to uniquely express $\Sigma$ as a function of $\Omega$
for a given $\dot M$. 

We should note here that although in the following we will mainly 
discuss disks with constant $\dot M$ our results are also directly 
applicable to disks in which $\dot M$ varies with distance. Indeed,
as long as one knows $\dot M$ at a particular distance (or $\Omega$)
in the disk equation (\ref{eq:rel1}) uniquely determines the 
value of $\Sigma$ in this location\footnote{Strictly speaking the 
relation $\dot M \sim\nu\Sigma$ used in deriving (\ref{eq:rel1}) 
is valid only in disks with smoothly varying $\Sigma$ (e.g. in disks 
with power law dependence of $\dot M$ on $r$); constant 
factor in this relation is in general different from $3\pi$.}. 
Thus, all our subsequent numerical
estimates would apply also to the case of non-constant $\dot M$ disks 
as long as $\dot M$ is specified at a location of interest.

We separately consider the cases of optically thick and optically 
thin gravitoturbulent disk. Before we do this we note that there
are two important transitions characterizing such 
disk. One is the 
\ba
\tau=1
\label{eq:tau1}
\ea
transition between the optically thick and optically thin regions. 
Another is the point at which $t_{cool}$ becomes 
comparable to $\Omega^{-1}$ and disk fragments. This transition is 
defined by condition 
\ba
\alpha_{GI}=\chi\sim 1,
\label{eq:frag}
\ea
where $\chi$ is the constant of order unity, its precise value 
has been determined by  Gammie (2001) and Rice \etal (2003, 2005) 
in a variety of circumstances. 

According to equations (\ref{eq:tau}) and 
(\ref{eq:alpha_GI}) each of these two relations sets a unique
constraint between $\Sigma$ and $\Omega$. However, if we demand
that both of them are fulfilled {\it simultaneously} (i.e. disk fragments 
exactly at the $\tau=1$ transition) then these relations 
hold only for specific values
of $\Sigma$ and $\Omega$, which 
we denote $\Sigma_f$ and $\Omega_f$. Equations (\ref{eq:tau}),
(\ref{eq:alpha_GI}), (\ref{eq:tau1}), and (\ref{eq:frag}) then yield
the following values of these parameters:
\ba
&& \Sigma_f\approx  \left[\left(\frac{\zeta\sigma}{\pi G Q_0\chi}\right)^{2\beta}
\left(\frac{\mu}{k_B}\right)^{\beta}\kappa_0^{-7}\right]^{(4\beta+7)^{-1}},
\label{eq:sig_f}\\
&& \Omega_f\approx \left[\left(\frac{\zeta\sigma}{\chi}\right)^{2\beta+1}
\frac{(\pi G Q_0)^{2\beta+6}}{\kappa_0^5}
\left(\frac{\mu}{k_B}\right)^{3\beta+4}\right]^{(4\beta+7)^{-1}}
\label{eq:omega_f}
\ea
(we set $f(\tau)\sim 1$ at $\tau=1$).
Through equations (\ref{eq:c_s}), (\ref{eq:T}), (\ref{eq:rel1}) 
$\Sigma_f$ and $\Omega_f$ (with $f(\tau)\sim 1$) also
determine fiducial values of the mass accretion rate $\dot M_f$,
sound speed $c_{s,f}$, and midplane temperature $T_f$:
\ba
&& \dot M_f\approx 3\pi\chi\nonumber\\
&& \times\left[\left(\frac{\zeta\sigma\kappa_0^2}
{\chi}\right)^{3}
(\pi G Q_0)^{4(\beta+1)}
\left(\frac{\mu}{k_B}\right)^{6(\beta+2)}\right]^{-(4\beta+7)^{-1}},
\label{eq:M_f}\\
&& c_{s,f}\approx \left[
\frac{\pi G Q_0\chi}{\zeta\sigma\kappa_0^{2}}
\left(\frac{\mu}{k_B}\right)^{-2(\beta+2)}\right]^{(4\beta+7)^{-1}},
\label{eq:cs_f}\\
&& T_f\approx \left[\left(
\frac{\pi G Q_0\chi}{\zeta\sigma\kappa_0^{2}}\right)^2
\frac{k_B}{\mu}\right]^{(4\beta+7)^{-1}},
\label{eq:T_f}
\ea
Note that all these fiducial quantities depend only on physical
constants and opacity parametrization. For
$\kappa$ given by equation (\ref{eq:dust_opacity}) 
we find the following values of these parameters (assuming 
$Q_0\approx 1$, $\chi,\zeta\sim 1$):
\ba
&& \Sigma_f\approx 15~\mbox{g}~\mbox{cm}^{-2},~~
c_{s,f}\approx 0.22~\mbox{km}~\mbox{s}^{-1},~~
T_f\approx 11.6~\mbox{K},
\nonumber\\
&& \Omega_f\approx 1.4\times 10^{-10}~\mbox{s}^{-1},~~
\dot M_f\approx 7\times 10^{-6}~\mbox{M}_{\odot}~\mbox{yr}^{-1}.
\label{eq:num_values}
\ea
The numerical value of $T_f$ conveniently falls into the regime 
of $\kappa$ dominated by icy dust grains.

For a given mass of a central object $M_\star$ angular frequency 
$\Omega_f$ determines a fiducial distance $r_f$ according
to the formula
\ba
r_f=\left(\frac{G M_\star}{\Omega_f^2}\right)^{1/3}\approx 130~\mbox{AU}
\left(\frac{M_\star}{M_\odot}\right)^{1/3}.
\label{eq:r_f}
\ea
In the case of a disk around $M_\star=10^6~M_\odot$ black hole one
finds $r_f\approx 0.06$ pc.

We now consider disk structure in the constant $\dot M$ case
separately for the optically thick and the optically thin regime, 
as well as the effects of external irradiation and the background 
viscosity in the disk.

\subsection{Optically thick case.}
\label{subsect:thick}

In the optically thick case $f(\tau)\approx\tau$. Plugging this into
equation (\ref{eq:rel1}) and using expression (\ref{eq:tau}) we find the 
following scalings:
\ba
&& \Sigma=\Sigma_f \left(\dot m\omega^{10-2\beta}\right)^{(7-2\beta)^{-1}}
=\Sigma_f \dot m^{1/3}\omega^{2},
\label{eq:Sig_thick}\\
&& T=T_f \left(\dot m^{2}\omega^{6}\right)^{(7-2\beta)^{-1}}=T_f 
\dot m^{2/3}\omega^{2},
\label{eq:T_thick}
\ea
where second equalities are for $\beta=2$ and we have defined 
the following dimensionless quantities:
\ba
\dot m\equiv\frac{\dot M}{\dot M_f},~~
\omega\equiv\frac{\Omega}{\Omega_f}.\nonumber
\ea
Using equations (\ref{eq:tau}) and (\ref{eq:alpha_GI}) we also find that 
in the optically thick regime
\ba
&& \alpha_{GI}=\chi\left(\dot m^{4-2\beta}\omega^{-9}
\right)^{(7-2\beta)^{-1}}=\chi\omega^{-3},
\label{eq:alpha_thick}\\
&& \tau=\left(\dot m^{2\beta+1}\omega^{10+4\beta}\right)^{(7-2\beta)^{-1}}=
\dot m^{5/3}\omega^6.
\label{eq:tau_thick}
\ea
According to equation (\ref{eq:tau_thick}) 
our assumption of $\tau>1$ is self-consistent only if 
\ba
\omega>\omega_1=\dot m^{-(2\beta+1)/(10+4\beta)} 
\label{eq:opt_tran}
\ea
(we assume $\beta<7/2$ as expected for dust opacity). Thus, gravitoturbulent
disk is going to be optically thick for all $\Omega>\Omega_f$ (or $\omega>1$)
if $\dot M>\dot M_f$ (or $\dot m>1$).
 
Note that equations (\ref{eq:Sig_thick})-(\ref{eq:tau_thick}) predict very 
rapid variation of disk properties with radius in the optically thick regime.
Indeed, for $\beta=2$ one finds $\Sigma,T\propto r^{-3}$, while 
$\tau\propto r^{-9}$. It is clear that even a moderate increase in $r$
would lead to the disk becoming optically thin. Also, 
$\alpha_{GI}\propto r^{9/2}$ (analogous scaling has been found by 
Goodman [2003] under different assumptions) and given this rapid variation 
it is obvious that an optically thick gravitoturbulent 
region can exist only within a limited range of radii, see
\S\S \ref{subsect:frag} \& \ref{subsect:bkgvisc}.

\subsection{Optically thin case.}
\label{subsect:thin}

In the optically thin regime we need to use $f(\tau)\approx \tau^{-1}$ in
equation (\ref{eq:rel1}) which combined with equation (\ref{eq:tau}) gives
the following scalings:
\ba
&& \Sigma=\Sigma_f \left(\dot m\omega^{10+2\beta}\right)^{(9+2\beta)^{-1}}
=\Sigma_f \dot m^{1/13}\omega^{14/13},
\label{eq:Sig_thin}\\
&& T=T_f \left(\dot m\omega\right)^{2/(9+2\beta)}=T_f 
\dot m^{2/13}\omega^{2/13},
\label{eq:T_thin}\\
&& \alpha_{GI}=\chi\left(\dot m^{2\beta+6}\omega^{-3}\right)^{(9+2\beta)^{-1}}
=\chi\dot m^{10/13}\omega^{-3/13},
\label{eq:alpha_thin}\\
&& \tau=\left(\dot m^{2\beta+1}\omega^{10+4\beta}\right)^{(9+2\beta)^{-1}}=
\dot m^{5/13}\omega^{18/13}.
\label{eq:tau_thin}
\ea
Assumption of an optically thin gravitoturbulent disk is valid provided 
that a condition opposite to (\ref{eq:opt_tran}) is satisfied. In
particular, $\tau<1$ for all $\Omega<\Omega_f$ (or $\omega<1$)
only if $\dot M<\dot M_f$ (or $\dot m<1$).

Properties of optically thin gravitoturbulent disks exhibit more 
moderate variation with $r$ than in the optically thick case. Indeed,
when $\tau<1$ and $\beta=2$ disk temperature and $\alpha_{GI}$ vary 
with distance quite slowly, $T\propto r^{-3/13}$ and 
$\alpha_{GI}\propto r^{9/26}$. 
As a result, optically thin gravitoturbulent regime can 
be realized in a rather extended region of the disk. 

As mentioned after equation (\ref{eq:rel1}) expressions 
(\ref{eq:Sig_thick})-(\ref{eq:tau_thick}) and 
(\ref{eq:Sig_thin})-(\ref{eq:tau_thin}) also apply to 
disks in which $\dot M$ varies with distance. In this
situation $\dot m=\dot m(\omega)$ and the dependence of various 
disk properties on $\omega$ can be directly obtained by plugging 
the expression for $\dot m(\omega)$ into these equations.

\subsection{External irradiation.}
\label{subsect:irrad}

According to equations (\ref{eq:T_thick}) and (\ref{eq:T_thin}) 
temperature of a gravitoturbulent disk decreases with $r$. 
At some point external irradiation 
becomes more important for the thermal balance
than the internal gravitoturbulent dissipation. Here we 
want to investigate a transition from a self-luminous disk (heated 
only by internal dissipation) to an irradiated regime. We assume 
that disk is illuminated by external radiation at temperature $T_0$. 
Given that most of the disk material is concentrated in its midplane
region the gravitational stability of the disk is going to be determined 
by the temperature $T$ near the midplane and the condition for marginal 
stability is still given by $Q\approx Q_0$ with $Q$ defined in 
equation (\ref{eq:Q}). Below we consider separately the optically 
thick and optically thin cases.

\subsubsection{Optically thick case.}
\label{subsubsect:irrad_thick}

Irradiated optically thick
disk appears as an extension of a self-luminous 
optically thick disk. A specific location at which such transition
occurs can be found by equating $T$ given by equation 
(\ref{eq:T_thick}) to $T_0$. Angular frequency corresponding to 
this transition is 
\ba
\omega_T=\dot m^{-1/3}\left(\frac{T_0}{T_f}\right)^{(7-2\beta)/6}.
\label{eq:omega_T_thick}
\ea
By construction $\tau$ at this location must be greater than 
unity, which with the 
aid of equation (\ref{eq:tau_thick}) yields the following 
constraint on the optically thick transition to an irradiated
disk:
\ba
\frac{T_0}{T_f}>\dot m^{(5+2\beta)^{-1}}.
\label{eq:tran_thick}
\ea

If $\tau\gtrsim 1$ then the midplane temperature of the disk is given 
by the solution of the vertical radiative transfer in diffusion 
approximation as
\ba
T^4\approx T_0^4+\eta\tau\frac{\dot M\Omega^2}{\sigma},
\label{eq:T_m}
\ea
where $\eta$ is a factor of order unity which absorbs our ignorance of how
the dissipation rate is distributed across the vertical thickness of the 
disk. Also, from energy conservation the temperature at the disk surface 
(photospheric temperature $T_{ph}$) is 
\ba
T_{ph}^4=T_0^4+\frac{3\dot M\Omega^2}{8\pi\sigma}.
\label{eq:T_ph}
\ea
These two relation allow us to distinguish three levels of the
importance of irradiation.

First, when 
\ba
\sigma T_0^4\lesssim \dot M\Omega^2
\label{eq:case1}
\ea
irradiation is so weak that it does not play any significant role even 
in the photosphere of the disk. This corresponds to the case of a 
self-luminous disk which was covered in \S \ref{sect:mdot}. 

Second,
when 
\ba
\dot M\Omega^2\lesssim\sigma T_0^4\lesssim \tau\dot M\Omega^2
\label{eq:case2}
\ea
external irradiation keeps disk surface temperature at the level of $T_0$
and creates a roughly isothermal gas layer underneath the surface. 
Nevertheless, the midplane temperature is still set predominantly by
the internal dissipation and the outgoing flux $\sim \dot M\Omega^2$
is still given by $F=\sigma T^4/\tau$, according to (\ref{eq:T_m}) and 
(\ref{eq:case2}). For that reason this irradiation regime also 
corresponds to the case of self-luminous disks explored in 
\S \ref{sect:mdot} with some minor corrections
having to do with the fact that the surface disk layers are hotter than
they would have been in the absence of irradiation.

Finally, when  
\ba
\tau\dot M\Omega^2\lesssim\sigma T_0^4
\label{eq:case3}
\ea
irradiation is so strong that even the midplane temperature reaches $T_0$
and the disk becomes vertically isothermal. This regime is 
different from the case of a self-luminous disk considered before since 
disk can no longer regulate its thermal state - its midplane temperature 
is fixed at $T\approx T_0$. Assuming that disk is gravitoturbulent
under strong irradiation we find from equation 
(\ref{eq:c_s}) that its surface density must scale linearly 
with $\Omega$:
\ba
\Sigma=\frac{c_{s,0}\Omega}{\pi G Q_0}\approx \Sigma_f
\left(\frac{T_0}{T_f}\right)^{1/2}\omega,
\label{eq:surf}
\ea
where $c_{s,0}$ is the sound speed corresponding to temperature $T_0$.
According to this result $\Sigma$ is independent of $\dot M$ in 
irradiated gravitoturbulent regions.

Under strong irradiation one can no longer use the expression 
(\ref{eq:t_cool}) to derive $\alpha_{GI}$. However, we know $\Sigma$
from (\ref{eq:surf}), so from the steady-state condition 
$\dot M=3\pi\nu\Sigma$ and definition $\nu=\alpha c_0^2/\Omega$ 
one finds that
\ba
\alpha_{GI}=\frac{Q_0}{3}\frac{G\dot M}{c_0^3}\approx \dot m
\left(\frac{T_0}{T_f}\right)^{-3/2}.
\label{eq:alph}
\ea
This result implies that $\alpha_{GI}$ is {\it constant} in the strongly 
irradiated part of the disk. Note that analogous result has been 
previously obtained in Goodman (2003). 

We can also calculate the run of the 
optical depth 
\ba
\tau=\frac{c_0\kappa_0T_0^\beta}{\pi G Q_0}\Omega\approx 
\left(\frac{T_0}{T_f}\right)^{\beta+1/2}\omega,
\label{eq:opt}
\ea
which shows that $\tau$ decreases as $r^{-3/2}$. This 
result combined with equation (\ref{eq:T_m}) also demonstrates 
that as $r$ increases the relative contribution of the internal
dissipation to the midplane temperature rapidly goes down since it is 
proportional to $\tau\Omega^2\propto\Omega^3$.

\subsubsection{Optically thin case.}
\label{subsubsect:irrad_thin}

Optically thin irradiated disk can emerge as an extension of 
an optically thin self-luminous disk, which happens when
$T$ given by equation (\ref{eq:T_thin}) becomes equal to $T_0$, 
or at 
\ba
\omega_T=\dot m^{-1}\left(\frac{T_0}{T_f}\right)^{(9+2\beta)/2}.
\label{eq:omega_T_thin}
\ea
Such an optically thin transition is possible when an inequality 
opposite to (\ref{eq:tran_thick}) is satisfied.

An irradiated optically thin disk can also appear as a continuation 
of the irradiated optically thick disk considered in \S 
\ref{subsect:thin}. According to equation 
(\ref{eq:opt}) this $\tau=1$ transition 
occurs at
\ba
\omega_T=\left(\frac{T_0}{T_f}\right)^{-(\beta+1/2)}.
\label{eq:omega_T_thin1}
\ea

Optically thin disk is roughly isothermal vertically and
its thermal balance requires 
\ba
T^4\approx T_0^4+\eta\frac{\dot M\Omega^2}{\tau\sigma},
\label{eq:th_bal}
\ea
which is different from equation (\ref{eq:T_m}) by a factor $\tau^{-1}$
in the second term on the right-hand side accounting for the 
inefficiency of radiative cooling and absorption in the optically 
thin disk. There are two obvious regimes to consider. First, when
\ba
\sigma T_0^4\lesssim\dot M\Omega^2/\tau,
\label{eq:case4}
\ea
irradiation does not affect disk properties and we go back to the case 
studied in \S \ref{subsect:thin}. Second, when the condition opposite 
to (\ref{eq:case4}) is satisfied irradiation sets the disk temperature.
In this case all results (except for the equation [\ref{eq:th_bal}]
different from [\ref{eq:T_m}]) obtained in the optically thick irradiated 
case -- equations (\ref{eq:surf})-(\ref{eq:opt}) -- remain valid 
since in deriving them we did not use 
any assumptions about the value of $\tau$.

The most important result regarding externally irradiated
constant $\dot M$ disks is that they can remain gravitoturbulent 
independent of their optical depth and that their effective viscosity 
$\alpha_{GI}$ remains constant. If the background viscosity 
does not dominate angular momentum transport at least in some 
self-luminous parts of the gravitoturbulent disk it will not 
dominate the transport in the irradiated part either since 
$\alpha_{GI}$ is constant there. Thus, torque needed for 
transporting mass through the disk must be due to the GI.

\subsection{Fragmentation limit.}
\label{subsect:frag}

When $t_{cool}$ becomes comparable to 
$\Omega^{-1}$ disk can no longer sustain the gravitoturbulence 
and has to fragment into bound objects (Gammie 2001). As mentioned 
before fragmentation condition $\Omega t_{cool}\lesssim 1$ can 
be recast in terms of the $\alpha_{GI}$ threshold 
according to equation (\ref{eq:frag}). This formulation now allows 
us to directly apply our results for $\alpha_{GI}$ derived in 
previous sections.

In particular, self-luminous optically thick gravitoturbulent 
disk starts fragmenting at  
\ba
\omega<\omega_{frag}=\dot m^{(4-2\beta)/9},
\label{eq:om_fr_thick}
\ea
which follows from demanding $\alpha_{GI}$ given by equation 
(\ref{eq:alpha_thick}) to be larger than $\chi$ -- the critical value 
of $\alpha$ needed for fragmentation. Radius $r_{frag}$ at which 
fragmentation first occurs is given by
$r_{frag}=r_f \dot m^{-4(2-\beta)/27}$ and is rather insensitive to
either $\beta$ or $\dot m$, so that fragmentation always occur 
not too far from $r_f$. It is rather interesting that for 
$\beta=2$, corresponding to the low temperature dust opacity 
the location of fragmentation edge in the optically thick limit 
is completely independent of $\dot m$: fragmentation
occurs exactly at $\Omega=\Omega_f$. This fact has been first noticed  
by Matzner \& Levin (2005). Clearly, constant $\dot M$ self-luminous 
gravitoturbulent disk cannot be fed by a source located outside 
$r_{frag}$. 

According to 
equation (\ref{eq:opt_tran}) $\omega_{frag}$ corresponds to the 
optically thick part of the disk only if $\dot m >1$. Thus, whenever
$\dot m>1$ the gravitoturbulent disk stays optically thick all the 
way to the fragmentation edge located inside of $r_f$. Then external 
disk feeding must necessarily occur interior to $r_f$.

In the optically thin case one finds from equation (\ref{eq:alpha_thin})
that 
\ba
\omega_{frag}=\dot m^{(2\beta+6)/3}.
\label{eq:om_fr_thin}
\ea
Fragmentation boundary lies in the optically thin part of the 
disk only if $\dot m<1$, in which case it is located outside
of the fiducial radius $r_f$. Fragmentation radius 
$r_{frag}=r_f \dot m^{-(2\beta+6)/9}$ is a rather sensitive 
function of $\dot m$:
for $\beta=2$ and $\dot M=10^{-7}$ M$_\odot$ 
yr$^{-1}\approx 0.014~\dot M_f$ (rather typical value of $\dot M$ in
protoplanetary disks) one finds $r_{frag}\approx 10^2 r_f$. Thus, in 
the optically thin regime fragmentation can be pushed out to 
large distances by reducing $\dot m$ (for the just used values 
of $\beta$ and $\dot M$ and
$M_\star=M_\odot$ one finds $r_{frag}\approx 10^4$ AU), but it still 
cannot be avoided if the disk is self-luminous. 

Everything we said before regarding fragmentation applied to self-luminous 
disks. If the disk is stable against fragmentation all the way to 
the point where its temperature is determined by external 
irradiation then in the irradiated region $\alpha_{GI}$ is constant 
and given by equation (\ref{eq:alph}). Since
$\alpha_{GI}$ must attain this value somewhere near the transition 
to irradiated regime and the disk is assumed to be non-fragmenting 
there (i.e. $\alpha_{GI}\lesssim \chi$ at $\omega\sim \omega_T$) we 
may conclude from equation (\ref{eq:alph}) that the disk is going 
to remain in a gravitoturbulent state stable 
against fragmentation as long as it is externally irradiated and 
\ba
\dot m\lesssim \chi\left(\frac{T_0}{T_f}\right)^{3/2}. 
\label{eq:stable}
\ea

This is a rather interesting conclusion since it implies that a 
sufficiently low $\dot M$ disk can in principle be stably fed by 
a source of mass located at infinity. In particular, for 
$T_0\approx 10^2$ K, when the low-temperature opacity 
(\ref{eq:dust_opacity}) still applies, one finds using estimates
(\ref{eq:num_values}) that $\dot M\lesssim 10^{-4}$ 
M$_\odot$ yr$^{-1}$ satisfies condition (\ref{eq:stable}) in 
which we set $\chi\sim 1$ for simplicity. Thus, disks obeying the condition 
(\ref{eq:stable}) can transfer mass at a constant rate from very 
large distances despite being gravitationally unstable, unlike the  
high $\dot M$ disks around quasars. Note that 
the criterion (\ref{eq:stable}) is independent of either the 
opacity behavior or the optical depth of the disk. A qualitatively 
similar conclusion about the stabilizing role of irradiation has 
been reached in Matzner \& Levin (2005) and Cai \etal (2008).  

The argument based on equation (\ref{eq:stable}) may not be 
without caveats. The value of $\alpha_{GI}$ given by equation 
(\ref{eq:alph}) is derived based on the cooling rate equal to 
the energy production rate $\sim \dot M\Omega^2$ due to the 
dissipation of gravitoturbulence. This rate is much lower than
the (external heating) rate $\sigma T_0^4 \times\min(1,\tau)$ at which 
the disk would cool if irradiation were suddenly switched off
or if the GI in irradiated disks were capable of producing 
surface density perturbations of order unity. One can easily show 
that $\Omega t_{cool}$ based on such a fast cooling rate does not
stay constant in irradiated part of the disk but steadily 
increases. The tricky question is the following: which cooling 
rate should determine the ability of the disk 
to fragment? This issue can be 
settled satisfactorily via careful numerical simulations of 
strongly irradiated 
gravitoturbulent disks, something that has not yet been done. 
But given that in accretion disks which are close to the steady state  
angular momentum transfer (determining $\dot M$) must be uniquely 
related to the energy dissipation rate we feel that it is 
more likely for the fragmentation condition to be determined by the 
criterion (\ref{eq:stable}) rather than by the much shorter cooling 
time set by the irradiation heating rate. Our subsequent consideration 
will be based on this assumption.

\subsection{Background viscosity.}
\label{subsect:bkgvisc}

If the angular momentum transport caused by gravitational 
torques becomes weak some other mechanisms may start providing
effective viscosity. Here we assume that in addition
to gravitational torques disk also possesses some background viscosity 
$\alpha_\nu$ due to e.g. the magneto-rotational instability (MRI). Normally
one expects $\alpha_\nu\ll 1$ so that this background viscosity would
become significant only when $\alpha_{GI}$ gets rather small. 

If, as expected for the dust opacity, $\alpha$ is between $0$ and $2$ then 
both in the optically thick and optically thin regimes $\alpha_{GI}$
decreases as $\omega$ increases. In other words, gravitoturbulent disk 
becomes less ``viscous'' as the distance to the center decreases. 
With this in mind we find that in the optically thick 
case the background viscosity would dominate over the gravitoturbulent  
torque (i.e. $\alpha_\nu\gtrsim\alpha_{GI}$) at 
\ba
\omega\gtrsim\omega_\nu\approx \left(\alpha_{\nu}^{2\beta-7}
\dot m^{4-2\beta}\right)^{1/9},~~~\tau>1.
\label{eq:om_nu_thick}
\ea
According to equation (\ref{eq:tau_thick}) this critical angular frequency 
corresponds to optically thick regime only if $\dot m\gtrsim
\alpha_{\nu}^{(10+4\beta)/(7+4\beta)}$. 

In the optically thin case 
background viscosity regulates disk at
\ba
\omega\gtrsim\omega_\nu\approx \left(\alpha_{\nu}^{-2\beta-9}
\dot m^{2\beta+6}\right)^{1/3},~~~\tau<1,
\label{eq:om_nu_thin}
\ea
and this $\omega_\nu$ corresponds to $\tau<1$ provided that
$\dot m\lesssim
\alpha_{\nu}^{(10+4\beta)/(7+4\beta)}$.

Inside the region where the background viscosity dominates 
(at $\omega>\omega_\nu$)  equations governing disk 
structure change. Previously, when considering the gravitoturbulent 
transport, we did not have a constraint on 
$\alpha$ but instead had a relationship between $T$ and $\Sigma$
in the form of equation (\ref{eq:T}), arising from the requirement of the 
marginal gravitational instability. However, with the background 
viscosity dominating the angular momentum transport $\alpha$ is constrained 
to be equal to $\alpha_{\nu}$, which leads to an overdetermined system of 
equations if we also try to keep the condition $Q=Q_0$. This contradiction
is naturally avoided by dropping the latter constraint, i.e. allowing
the disk not to be marginally gravitationally unstable when 
$\alpha=\alpha_\nu$. 

In Appendix \ref{app1} we present $\Sigma$, $T$, and $\tau$
behaviors in accretion disk with the dominant background viscosity 
$\alpha_\nu$ and opacity in the form (\ref{eq:opacity}).
If we now use these expressions [equations (\ref{eq:Sig_eq_thick}), 
(\ref{eq:T_eq_thick}), (\ref{eq:Sig_eq_thin}), and  
(\ref{eq:T_eq_thin})] to calculate Toomre $Q$ in the self-luminous viscous
part of the disk at $\omega>\omega_\nu$ we find that
\ba
Q/Q_0=(\omega/\omega_\nu)^{9/(10-2\beta)},~~~\tau>1
\label{eq:Qthick}
\ea
in the optically thick case and 
\ba
Q/Q_0=(\omega/\omega_\nu)^{3/(6+2\beta)},~~~\tau<1
\label{eq:Qthin}
\ea
in the optically thin case. Apparently, for any reasonable dust opacity
behavior $Q$ starts to deviate from its marginal stability value $Q_0$
towards higher values at the transition from the gravitoturbulent to 
viscous regime (at $\omega=\omega_\nu$). Thus, accretion disk cannot be 
gravitationally unstable if its angular momentum transport is not dominated 
by gravitational torques but is rather due to some other form of 
effective viscosity. This consideration demonstrates explicitly 
how the transition between the gravitoturbulent and viscous parts of the 
disk occur.

Note that $\alpha_{GI}$ given by equation (\ref{eq:alph}) is independent
of $\omega$. This implies that if 
\ba
\dot m\lesssim \alpha_\nu\left(\frac{T_0}{T_f}\right)^{3/2},
\label{eq:dotm_visc_limit}
\ea
accretion disk remains viscous everywhere and does not transition into the 
gravitoturbulent state at all. Thus, at low enough
$\dot m$ a gravitationally stable viscous disk can transport 
matter all the way to the central object from infinitely large distances.
Condition (\ref{eq:dotm_visc_limit}) is more restrictive than the
constraint (\ref{eq:stable}) which delineates regime in which the 
gravitoturbulent constant $\dot M$ disk can transfer material from 
infinitely large distances without fragmentation.

\section{Discussion.}
\label{sect:disc}

Our results derived in \S \ref{sect:mdot} apply to a variety of
situations in which gravitoturbulent disks can exist: they can be
optically thick or thin, fragmenting or not. Here we classify 
different states in which accretion disks can be found according to
the values of $\dot M$, background viscosity $\alpha_\nu$ and 
irradiation temperature $T_0$. We also describe applications of 
these results to real astrophysical systems and discuss their
connection with the work of others.

\subsection{Separation of different regimes.}
\label{subsect:regimes}

As we saw in previous sections different parts of accretion disks 
can be characterized by different states --- gravitoturbulent, 
viscous, irradiated, etc. Transitions between these regimes depend
on $\dot M$, $T_0$, 
and $\alpha_\nu$. Our results allow us to 
single out three important cases for the different transition 
topologies. 

\begin{figure}
\plotone{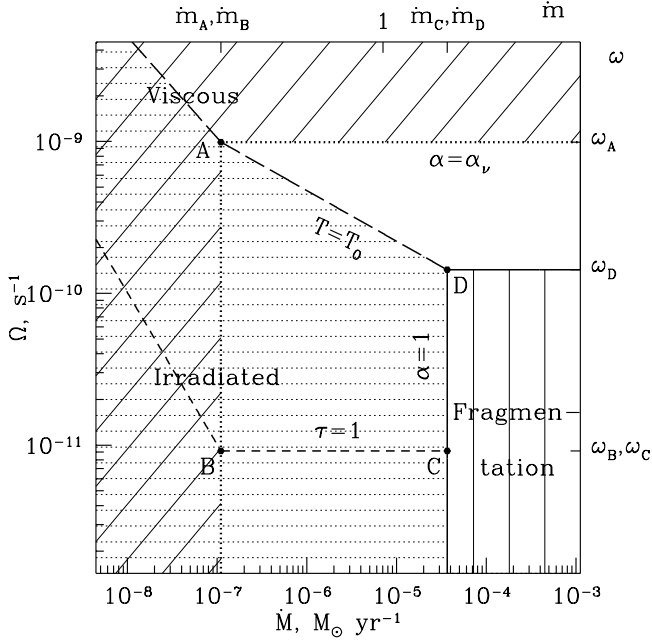}
\caption{
Plot of different possible states in which an accretion disk can 
be found as a function of mass accretion rate $\dot M$ and angular 
frequency $\Omega$ (corresponding dimensionless quantities 
$\dot m=\dot M/\dot M_f$ and $\omega=\Omega/\Omega_f$ are shown on the
upper and right axes). Disk has a non-zero background viscosity 
$\alpha_\nu=0.003$ and is externally irradiated at a temperature 
$T_0=3T_f\approx 35$ K, corresponding to condition (\ref{eq:CaseA}).
Shading indicates viscous regions ({\it slant solid shading}), 
irradiated regions ({\it horizontal dotted shading}), and region 
where the disk must fragment ({\it vertical solid shading}). 
Gravitoturbulent, self-luminous region is unshaded. Different curves
separate regions with distinct physical conditions and are marked on the 
plot (see text for more details). Coordinates of the points where 
these curves cross in the $(\dot m,\omega)$ coordinates are 
given by expressions (\ref{eq:CaseA_points}). 
\label{fig:CaseA}}
\end{figure}

When external irradiation is strong, 
\ba
T_0/T_f>1,
\label{eq:CaseA}
\ea
the phase space 
of possible regimes can be represented by Figure \ref{fig:CaseA}. 
Given that in the case of opacity dominated by cold dust $T_f$ 
is just slightly higher than $10$ K (see eq. [\ref{eq:num_values}])
it is clear that the condition (\ref{eq:CaseA}) should apply to 
virtually all accretion 
disks in the Universe as even the lowest measured temperatures 
encountered in some dense molecular cores are not very different from 
$10$ K (Di Francesco \etal 2007). 

Using Figure \ref{fig:CaseA} one can examine different regimes 
(indicated by shading) which are relevant for a given $\dot M$. 
Curves\footnote{Equations for these curves
can be found in \S\S \ref{subsect:thick}-\ref{subsect:bkgvisc} 
and Appendix \ref{app1}.} 
separating different regimes correspond to various critical 
transitions: $\alpha=\alpha_\nu$ (dotted lines) implies a transition 
from viscous (above and to the left of this curve, slant solid 
shading) to gravitoturbulent state, $\alpha=1$ (solid lines) 
describes the onset of fragmentation (below and to the right 
of this curve, vertical solid shading), 
$T=T_0$ is a (long-dashed) line below which 
external irradiation starts to dominate disk structure 
(region with horizontal dotted shading), $\tau=1$ curve (short-dashed) shows 
a transition from an optically thick (above 
this curve) to an optically thin disk. Coordinates (in phase space 
$\dot m,\omega$, upper and right axes) of the points 
A, B, C, D where these curves cross are given by expressions 
(\ref{eq:CaseA_points}) in Appendix \ref{app2}.
Unshaded region corresponds to a self-luminous, gravitoturbulent disk.  
A constant $\dot M$ disk corresponds to
a straight vertical line in Figure \ref{fig:CaseA}  cutting 
through different regimes.

The numerical values on the left and lower axes of this Figure 
correspond to a particular 
choice of $\alpha_\nu=0.003$, $T_0/T_f=3$ ($T_0=35$ K) and 
opacity in the form (\ref{eq:dust_opacity}).
As described in \S \ref{subsect:bkgvisc} at the very low 
$\dot M\lesssim 10^{-7}$ M$_\odot$ yr$^{-1}$ disk is always
viscous and gravitationally stable, even at arbitrarily large distances 
from the central object. Above this value of $\dot M$ the disk must be  
gravitoturbulent within some range of distances. For 
$10^{-7}$ M$_\odot$ yr$^{-1}\lesssim\dot M\lesssim 4\times 10^{-5}$ 
M$_\odot$ yr$^{-1}$ gravitoturbulent region extends from $\Omega\sim 
10^{-10}-10^{-9}$ s$^{-1}$ all the way to very large distances 
(where $\Omega\to 0$). However, for $\dot M$ above $\dot M\gtrsim 
4\times 10^{-5}$ M$_\odot$ yr$^{-1}$ gravitoturbulent disk fragments 
at $\Omega\approx 1.4\times 10^{-10}$ s$^{-1}$ (in the optically thick 
case with opacity characterized by $\beta=2$ 
fragmentation occurs exactly at $\Omega_f$, see \S \ref{subsect:frag}) 
so that a constant $\dot M$ disk can be maintained only interior 
to this point. For any $\dot M\gtrsim 10^{-7}$ M$_\odot$ yr$^{-1}$
the gravitoturbulent angular momentum transport becomes 
weak at large enough $\Omega$ so that the background viscosity starts 
to determine disk properties at small radii 
(upper part of the plot, at $\Omega>10^{-9}$ s$^{-1}$ in this 
particular case). 

\begin{figure}
\plotone{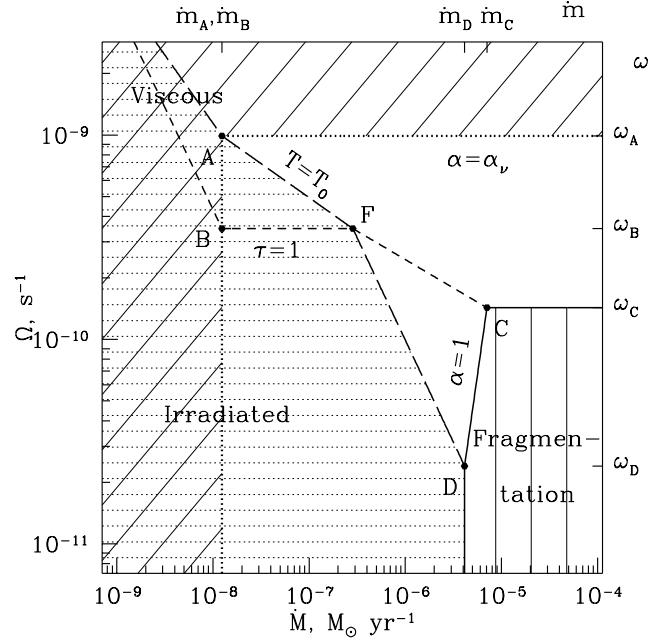}
\caption{
Same as Figure \ref{fig:CaseA} but for $\alpha_\nu=0.003$ and 
$T_0=0.7T_f\approx 8$ K, corresponding to the condition 
(\ref{eq:CaseB}). Coordinates of the points where 
different curves cross are 
given by the expressions (\ref{eq:CaseB_points}). 
\label{fig:CaseB}}
\end{figure}

As an example, let us use Figure \ref{fig:CaseA} to figure out where 
the transitions between different regimes occur in
a disk with $\dot M=10^{-5}$ M$_\odot$ yr$^{-1}$, 
$\alpha_\nu=0.003$, $T_0/T_f=3$ and opacity in the form 
(\ref{eq:dust_opacity}). At $\Omega>10^{-9}$ s$^{-1}$ disk is 
gravitationally stable and self-luminous, angular momentum is 
transported through the disk by background viscosity. 
For $2\times 10^{-10}$ s$^{-1}$ $<\Omega<10^{-9}$ 
s$^{-1}$ disk is gravitoturbulent and self-luminous. 
For $\Omega<2\times 10^{-10}$ s$^{-1}$ extending to infinity disk 
is gravitoturbulent and externally irradiated. Disk is optically 
thick for all $\Omega\gtrsim 10^{-11}$ s$^{-1}$ and optically thin
for smaller $\Omega$. This example clearly
shows the complexity of possible states in which a given accretion disk 
can be found at different distances from the central object.

Topology of the phase space of possible disk states changes when the
temperature of external radiation field becomes smaller than $T_f$. 
This situation is not very typical (remember that the CMB temperature 
in the present day Universe which would set a temperature floor is not 
too far from $T_f$) for astrophysical objects although temperature 
at the level $6-7$ K has been found in centers of some
dense molecular cores well shielded from external starlight, so 
that situations in which $T_0<T_f$ can be relevant for very young 
protoplanetary disks (Class 0 objects without central heating source) 
forming inside very dense cores. 

For completeness we display the different 
topologies of disk states possible for $T_0<T_f$. There are two 
cases. First, for
\ba
\alpha_\nu^{2/(7+4\beta)}<T_0/T_f<1
\label{eq:CaseB}
\ea
the separation of different regimes in the $\dot M, \Omega$ space is 
represented by Figure \ref{fig:CaseB}. One can see from this Figure
that the range of $\dot M$ in which there is a gravitoturbulent region 
extending to infinity shifts to lower $\dot M$ (Figure \ref{fig:CaseB} 
is drawn for the same $\alpha_\nu=0.003$ as Figure \ref{fig:CaseA} but
for $T_0=0.7T_f\approx 8$ K).

\begin{figure}
\plotone{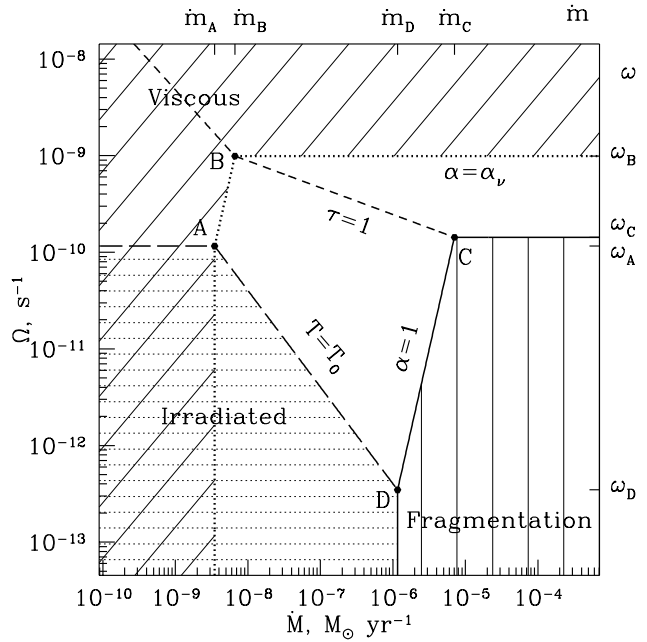}
\caption{
Same as Figure \ref{fig:CaseA} but for $\alpha_\nu=0.003$ and 
$T_0=0.3T_f\approx 3.5$ K, corresponding to the condition 
(\ref{eq:CaseC}). Coordinates of the points where 
different curves cross are 
given by the expressions (\ref{eq:CaseC_points}). 
\label{fig:CaseC}}
\end{figure}

Second, for very low irradiation temperature
\ba
T_0/T_f<\alpha_\nu^{2/(7+4\beta)}
\label{eq:CaseC}
\ea
the  separation of different regimes is shown in Figure \ref{fig:CaseC}.
This plot again assumes $\alpha_\nu=0.003$ while 
$T_0=0.3T_f\approx 3.5$ K. Such low temperature is hardly achievable 
in reality as it is just slightly higher than the current CMB 
temperature. Note that the topology of different regimes in Figure
\ref{fig:CaseB} is intermediate between the cases shown in Figures
\ref{fig:CaseA} and \ref{fig:CaseC} --- it looks like the former
at low $\dot M$ and like the latter at high $\dot M$.

Despite some differences in details of how exactly the 
transition from viscous to gravitoturbulent, to irradiated states 
occurs at intermediate values of $\dot M$, 
Figures \ref{fig:CaseA}-\ref{fig:CaseC} 
exhibit four invariant properties of accretion disks: (1) disk is 
dominated by background viscosity for all $\Omega$ at very low 
$\dot M$ [for $\dot m<\alpha_\nu (T_0/T_f)^{3/2}$, 
see Appendix \ref{app2}] and non-zero $T_0$, (2) for 
all $\dot M$ disk is dominated 
by background viscosity at high enough $\Omega$, (3) at intermediate 
values of $\dot M$ [for $\alpha_\nu (T_0/T_f)^{3/2}<\dot m<(T_0/T_f)^{3/2}$] 
disk possesses a gravitoturbulent, externally 
irradiated region that extends to arbitrarily large distances, (4)
at large values of $\dot M$ [for 
$\dot m>(T_0/T_f)^{3/2}$] disk has a gravitoturbulent region within 
a finite range of distances but must inevitably fragment at some 
large distance. 
Similar qualitative conclusions hold also 
for opacity behaviors different from (\ref{eq:dust_opacity}).

We should note here that calculations presented in this section rely 
on our use of viscosity (\ref{eq:opacity})-(\ref{eq:dust_opacity}) 
throughout the whole region of $\dot M,\Omega$ phase space that we 
consider. In reality, at high $\dot M$ and $\Omega$ disk temperature
should exceed $10^2$ K at which point icy grains sublimate leaving
metal-silicate grains as a source of opacity. This latter opacity 
source while still being in the form (\ref{eq:opacity}), is 
characterized by smaller values of $\kappa$ at the same temperature and
$\beta\approx 1/2$. This is likely to quantitatively (but not 
qualitatively) affect results presented in Figures 
\ref{fig:CaseA}-\ref{fig:CaseC} at high $\dot M$ and $\Omega$.
To not complicate things further here we do not attempt to self-consistently 
describe transitions between different opacity regimes 
but rather display a qualitative picture for a single opacity law.

\subsection{Applications.}
\label{subsect:apps}

Our results can be applied to understanding the properties of the 
outer, cold parts of realistic accretion disks. In particular, we
address three important issues.

First, a possibility of giant planet formation by GI in protoplanetary 
disks has been discussed since Cameron (1978). In this context it
is interesting to ask under which conditions a constant $\dot M$ 
protoplanetary disk would be prone to fragmentation into 
gravitationally bound, self-gravitating objects. Our results 
described in \S \ref{subsect:regimes} can directly address this 
issue. Indeed, conditions used in producing Figure \ref{fig:CaseA}, 
namely $\alpha=0.003$ and $T_0\approx 35$ K are quite typical for 
the outer regions of protoplanetary disks, beyond $\sim 100$ AU
from the central star. One one hand, the disk surface density there 
is low enough (see below) for the cosmic ray ionization to stimulate
MRI operation which gives rise to $\alpha_\nu$ at the level of 
$\sim 10^{-3}-10^{-2}$. On the other hand, outer regions of protoplanetary
disks are warmed up by radiation of either the parent star or the
neighboring stars at the level of several tens of K. Thus, the situation 
represented in Figure \ref{fig:CaseA} can be directly used for 
understanding the properties of external parts of protoplanetary 
disks.

What is obvious from this Figure is that giant planet formation by
gravitational instability in constant $\dot M$ disks can take place 
only beyond $\approx 120$ AU which is the distance from a $1$ M$_\odot$ 
star at which $\Omega=\Omega_f$ -- remember that in the optically 
thick\footnote{That the disk with the assumed values of $\alpha_\nu$ and 
$T_0$ is optically thick at fragmentation boundary is evident from Figure 
\ref{fig:CaseA}.} gravitoturbulent disks fragmentation occurs 
at this specific value of $\Omega$ (Matzner \& Levin 2005), see 
equation (\ref{eq:om_fr_thick}). Thus, planets produced by gravitational
instability should be born far from their parent stars although 
one cannot exclude their subsequent migration to shorter periods.  

Another obvious constraint on planet formation that follows from
Figure \ref{fig:CaseA} is that $\dot M$ must be pretty high at the 
location where disk fragments and planets form. Indeed, one can easily 
see that fragmentation is possible only if $\dot M$ locally exceeds $10^{-5}$
M$_\odot$ yr$^{-1}$. At lower $\dot M$ disk maintains itself in 
a gravitoturbulent state (or even being kept gravitationally stable 
by its own background viscosity at very low $\dot M$) without 
fragmentation even very far from the star. 

Accretion rates in excess of $10^{-5}$ M$_\odot$ yr$^{-1}$ are 
atypical for mature T Tauri disks (Gullbring \etal 1998). 
However, they may have been present at 
the very earliest stages of star and disk formation when the material
from collapsing protostellar envelope rains down onto the disk at a very
high rate, possibly exceeding  $10^{-5}$ M$_\odot$ yr$^{-1}$ in 
some locations. Such disks are likely not to have $\dot M$ constant
through their whole extent but as we discussed in \S \ref{sect:mdot} 
our results are still applicable\footnote{A disk with $\dot M$ varying 
with distance would not correspond to a straight vertical line in 
Figures \ref{fig:CaseA}-\ref{fig:CaseC} like a constant $\dot M$ disk 
would but must follow a more complicated path determined by a specific 
dependence of $\dot M$ on $r$ (or, alternatively, $\Omega$).} 
even in this more complicated case as long as $\dot M$ is specified
locally. 

Second practical issue that we are going to address has to do with the 
feeding of supermassive black holes in centers of galaxies. 
It has been known for a long time that the outer parts of quasar 
disks must be gravitationally unstable which was always raising 
a question of how gas is transported to the black hole from 
large distances. Our results demonstrate that as long as $\dot M$ 
is not very high the GI is not going to impede 
mass transfer through the disk since for low enough $\dot M$, namely
for $\dot M\lesssim M_f(T_0/T_f)^{3/2}$, see equation (\ref{eq:stable}), 
a gravitationally
unstable disk can persist in a gravitoturbulent state at arbitrarily 
large distances from the central object. How high $\dot M$ can be
carried through a gravitoturbulent disk globally thus depends only 
on the level of external irradiation.

Radiation fields in galactic nuclei due to circumnuclear stars 
are expected to be quite intense giving rise to $T_0$ at the level of tens 
to hundreds of K. Assuming $T_0=100$ K (which is a radiation field 
slightly more intense than that expected in the Galactic Center) we 
find that gravitoturbulent disk can transport mass from infinity to 
the black hole as long as $\dot M\lesssim 10^3\dot M_f\approx 10^{-2}$
M$_\odot$ yr$^{-1}$. This is about $10\%$ of the Eddington rate 
(for radiative efficiency of $10\%$) for the $4\times 10^{6}$ M$_\odot$
black hole in the center of our Galaxy, which is quite significant given 
that the Bondi accretion rate of this object is $\lesssim 10^{-5}$
M$_\odot$ yr$^{-1}$ (Baganoff \etal 2003).
Thus, irradiated gravitoturbulent accretion disks provide a natural 
way of continuous feeding at least some (not too massive) central black 
holes at reasonable rates by gas transported from very large distances.  

However, it is also clear from our results and Figure \ref{fig:CaseA}
that accretion disks around more massive black holes ($>10^6$ M$_\odot$) 
consuming mass at rates close to Eddington cannot remain 
gravitoturbulent out to very large distances --- for any reasonable 
level of external irradiation disk must fragment at some point, and the 
most distant possible location of the fragmentation boundary  
corresponds to $\Omega=\Omega_f\approx 1.4\times 10^{-10}$ s$^{-1}$ (for the
case of dust opacity in the form [\ref{eq:dust_opacity}]). Transport of
gas from beyond this distance is still an open issue discussed by Goodman 
(2003).

Finally, we briefly discuss the origin of young stellar disk 
around a supermassive black hole in the center of our Galaxy. Inner
parsec of the Galaxy is known (Paumard \etal 
2006; Lu \etal 2009) to contain at least one disk of young ($6\pm 2$ 
Myr old) massive stars spread between $0.04$ pc and $0.5$ pc. 
To explain formation of these stars so close to the black hole where 
they are subject to action of its strong tidal field a fragmentation 
of a gravitationally
unstable disk has been proposed (Levin \& Beloborodov 2003; Levin 2007).
Such an event can in principle happen both in a (quasi-)steady state disk
like the one we considered in this work or in a short-lived promptly
fragmenting (on a dynamical time scale) 
disk-like structure that may arise as a result 
of molecular cloud collision with the
black hole (Wardle \& Yusef-Zadeh 2008; Bonnell \& Rice 2008; 
Hobbs \& Nayakshin 2008). Here we try to constrain the first 
possibility. We will assume that disk was illuminated 
by surrounding stars which kept $T_0$ at the level of tens of K so that 
its opacity law was given by equation (\ref{eq:dust_opacity}). Distance
$r_f$ at which fragmentation would occur in an optically thick disk 
around the $4\times 10^6$ M$_\odot$ black hole is $\approx 0.1$ pc which
is within the span of the observed stellar disk.

For a long-lived disk to start fragmenting a variability of some of its
properties must be taking place. One possibility is an increase of 
$\dot M$ which can bring a gravitoturbulent irradiated disk extending
out to large distances across a fragmentation threshold (see Figure 
\ref{fig:CaseA}). But then the outer regions of the disk 
(at $r\gg r_f$) where 
$\dot M$ has increased would immediately fragment and it is not at 
all obvious that mass would be transferred inward 
increasing $\dot M$ at small radii (at $r\sim r_f$) where stellar 
disks are observed. In principle, disk could be not an accretion but a
spreading (Pringle 1991) disk formed as a result of a dense molecular 
cloud disruption at very small distances. However, in this case an 
{\it increase} of $\dot M$ is unlikely while a decrease of $\dot M$
typical for spreading disks would only stabilize the
disk against gravitational fragmentation.

Another (probably less likely) possibility is a reduction of external  
irradiation which can cause fragmentation of even a constant in time
$\dot M$ disk as soon as the condition (\ref{eq:stable}) gets violated. 
The problem with this scenario is that it may then be difficult to explain 
the existence of stars at $0.04$ pc which is significantly closer to
the black hole than 
the minimum radius $r_f$ at which fragmentation occurs in a cold 
disk. In principle, $\Omega_f$ can be increased by lowering disk 
metallicity which affects $\kappa_0$. However, according to
equations (\ref{eq:omega_f}) \& (\ref{eq:r_f}) moving $r_f$ from 0.1 pc
to 0.04 pc would require reducing $\kappa_0$ by a factor of $60$ compared 
to the value in (\ref{eq:dust_opacity}), implying extremely sub-solar
metallicity in the disk. 

More generally, star formation in a 
nearly-Keplerian disk does not naturally explain rather significant
eccentricities of disk stars (Bartko \etal 2008) and the possible 
presence of a second disk component. Thus, it seems unlikely (although 
not completely impossible) that stellar disks were formed by gravitational 
fragmentation of a long-lived gaseous disk. Scenario of a prompt 
fragmentation of a tidally disrupted molecular cloud advanced by 
Wardle \& Yusef-Zadeh (2008), Bonnell \& Rice (2008) and  
Hobbs \& Nayakshin (2008) presents a more attractive possibility.

\subsection{Comparison with previous studies.}
\label{subsect:compare}

The first investigation of self-gravitating accretion disks 
has been undertaken by Paczynski 
(1978a), later followed by Paczyncki (1978b) and Kozlowski \etal 
(1979). These early numerical modelling efforts concentrated on
studying hot quasar disks around the supermassive black holes with 
accretion rates close to the Eddington rate. Some 
of the important ingredients of these models have been the
inclusion of the radiation pressure (neglected in our case) and use
of high-$T$ opacities, 
which makes comparison of these calculations to the results 
of our study rather difficult. 

Another investigation of a quasi-viscous evolution of a 
self-gravitating disk driven by gravitational torques has been 
done by Lin \& Pringle (1987). In their work, based on rather general 
arguments, a specific model for the viscosity due to the disk 
self-gravity has been assumed, namely $\nu\propto \Sigma^2r^6\Omega$
(or $\alpha\sim Q^{-2}> 1$). 
The correct prescription (\ref{eq:alpha}) is in general quite different
from this naive anzatz precluding direct comparisons with our results. 
Also, the assumption $Q\sim 1$ has been relaxed in 
Lin \& Pringle (1987) allowing $Q$ to drop significantly below unity. 

More recently Goodman (2003) has analytically investigated properties of 
$Q\approx 1$ regions of constant $\dot M$ quasar disks, again taking 
into account radiation pressure. This study has a 
lot in common with our work with the major difference being that 
Goodman (2003) left $\alpha$ to be a free parameter while we 
self-consistently calculate its value using a prescription 
(\ref{eq:alpha}). Some of the results derived in Goodman (2003)
have been retrieved in our study.

Subsequently, Rafikov (2005, 2007) has looked at the properties of 
gravitationally unstable disks which are capable of forming giant planets
by direct fragmentation. Such disks must simultaneously fulfill 
two constraints: $Q=Q_0$ and $\alpha_{GI}=\chi$. 
These assumptions were used to set stringent constraints
on the properties of disks that are able to form planets by 
GI. Clearly, such disks do not have constant 
$\dot M$ in general. Our current assumptions are
different in that we assume only $Q=Q_0$ and fix $\dot M$ at some value
(not necessarily constant, see the discussion after equation 
[\ref{eq:rel1}]) at every point in the disk, but $\alpha_{GI}$ is 
then calculated self-consistently. Matzner \& Levin (2005) and Levin 
(2007) have also studied cold 
$Q=1$ disks with opacity dominated by dust specifically looking at the 
conditions necessary for fragmentation. We
successfully reproduce some of their results such as the location 
of the fragmentation boundary in the optically thick regime and 
the importance of irradiation for stabilizing the gravitoturbulent disk
against fragmentation at large distances. 

Finally, Terquem (2008) \& Zhi \etal(2008) have constructed global 
numerical models of protoplanetary disks accounting for the possibility 
of GI in some parts of the disk. These studies 
pay special attention to the presence of the so called 
``dead zones'' (Gammie 1996) --- disk regions where MRI cannot 
operate because of low ionization. Our analytical calculations do not account 
for the existence of such regions, nevertheless they provide good 
foundation for understanding numerical results of Terquem (2008) 
\& Zhi \etal(2008) in the outermost regions of their disks.

\section{Conclusions.}
\label{sect:concl}

We have explored the properties of marginally gravitationally 
unstable accretion disks using a realistic prescription for
the angular momentum transfer driven by the gravitational torques. 
We self-consistently derived scalings of important disk variables such as 
surface density and temperature in both the optically thick and 
thin regimes. We also accounted for the possibility of
disk having some background viscosity, e.g. due to MRI, 
and demonstrated that in this case a gravitoturbulent disk inevitably 
switches to an ordinary viscous disk at small radii. Another important
ingredient of our study is the inclusion of possible external 
irradiation of the disk (e.g. by central object or nearby stars).
We have demonstrated that for low enough $\dot M$ external 
irradiation helps to stabilize the disk against fragmentation all
the way to infinity providing means of mass transport to the 
central object from very large distances. At extremely low $\dot M$
disks have been shown to never become gravitationally unstable
because of background viscosity.
At high $\dot M$ (the exact threshold depends on the irradiation
temperature) fragmentation of the disk is inevitable at large 
distances. Results of this work apply to our understanding 
of the possibility of giant planet formation by GI and star formation 
in the Galactic Center, and to the 
problem of feeding the quasars.


\acknowledgements 

I am grateful to Jeremy Goodman for useful discussions. 
The financial support for this work is provided by the Sloan Foundation
and NASA grant NNX08AH87G.

\appendix

\section{Properties of viscous accretion disks.}\label{app1}

When the angular momentum transport in the disk is dominated by
some internal source of viscosity characterized by parameter 
$\alpha_\nu$ one finds that in the 
optically thick case
\ba
&& \Sigma_\nu\approx\left[\frac{\sigma}{\kappa_0}
\frac{\Omega^{2-\beta}\dot M^{3-\beta}}{\alpha_\nu^{4-\beta}}
\left(\frac{k_B}{\mu}\right)^{\beta-4}\right]^{(5-\beta)^{-1}}\approx 
\Sigma_f\left(\frac{\omega^{2-\beta}\dot m^{3-\beta}}
{\alpha_\nu^{4-\beta}}\right)^{(5-\beta)^{-1}},
\label{eq:Sig_eq_thick}\\
&& T_\nu\approx\left[\Omega^3\frac{\kappa_0\dot M^2}
{\sigma\alpha_\nu}\frac{\mu}{k_B}
\right]^{(5-\beta)^{-1}}\approx 
T_f\left(\frac{\omega^3\dot m^2}
{\alpha_\nu}\right)^{(5-\beta)^{-1}},~~~\tau>1,
\label{eq:T_eq_thick}
\ea
where we have dropped constant numerical factors of order unity.
With these expressions one finds 
\ba
\tau=\left(\frac{\omega^{2\beta+2}\dot m^{\beta+3}}
{\alpha_\nu^4}\right)^{(5-\beta)^{-1}},
\label{eq:tau_eq_thick}
\ea
meaning that viscous disk is optically thick only when 
\ba
\omega>\left(\frac{\alpha_\nu^4}{\dot m^{\beta+3}}
\right)^{(2+2\beta)^{-1}}.
\label{eq:tau_1_omega}
\ea

When the condition (\ref{eq:tau_1_omega}) is violated 
disk becomes optically thin and we find
\ba
&& \Sigma_\nu\approx\left[\sigma\kappa_0
\frac{\Omega^{2+\beta}\dot M^{3+\beta}}{\alpha_\nu^{4+\beta}}
\left(\frac{\mu}{k_B}\right)^{4+\beta}\right]^{(3+\beta)^{-1}}
\approx \Sigma_f \dot m\left(\frac{\omega^{2+\beta}}
{\alpha_\nu^{4+\beta}}\right)^{(3+\beta)^{-1}},
\label{eq:Sig_eq_thin}\\
&& T_\nu\approx\left[\Omega\frac{\alpha_\nu}{\sigma\kappa_0}
\frac{k_B}{\mu}\right]^{(3+\beta)^{-1}}\approx 
T_f \left(\alpha_\nu\omega\right)^{(3+\beta)^{-1}},
\label{eq:T_eq_thin}\\
&& \tau=\dot m\left(\frac{\omega^{1+\beta}}
{\alpha_\nu^2}\right)^{2/(3+\beta)},~~~\tau<1.
\label{eq:tau_eq_thin}
\ea

Derivation of equations (\ref{eq:Sig_eq_thick})-(\ref{eq:tau_eq_thin}) 
assumes that it is the internal viscous dissipation in the disk that
sets its midplane temperature. However, analogous to the case studied
in \S \ref{subsect:irrad} one can consider a possibility that the disk 
temperature is set by external irradiation at the level of $T=T_0$ 
even when its angular momentum transport has non-gravitational nature.
From equation (\ref{eq:T_eq_thick}) we find that an optically thick 
self-luminous viscous disk changes to an externally irradiated disk 
at a radius where 
\ba
\omega=\omega_T=\left(\frac{\alpha_\nu}{\dot m^2}\right)^{1/3}
\left(\frac{T_0}{T_f}\right)^{(5-3)/3}. 
\label{eq:T_thick_omega}
\ea
Analogously, equation (\ref{eq:T_eq_thin}) predicts that an optically thin 
transition from a self-luminous to an externally irradiated disk 
occurs at the point where 
\ba
\omega=\omega_T=\alpha_\nu^{-1}
\left(\frac{T_0}{T_f}\right)^{(3+\beta)}. 
\label{eq:T_thin_omega}
\ea

In those regions where the disk temperature is fixed at the constant 
level $T_0$ one can easily show that
\ba
&& \Sigma=\frac{\dot m\omega}{\alpha_\nu}
\frac{T_0}{T_f},
\label{eq:Sig_eq_irrad}\\
&& \tau=\frac{\dot m\omega}{\alpha_\nu}
\left(\frac{T_0}{T_f}\right)^{(\beta-1)}.
\label{eq:tau_eq_irrad}\\
&& Q/Q_0=\frac{\alpha_\nu}{\dot m}
\left(\frac{T_0}{T_f}\right)^{3/2}.
\label{eq:Q_visc_irrad}
\ea
Last equation implies that Toomre $Q$ is constant in the irradiated 
viscous part of the constant $\dot M$ disk. That $Q>Q_0$ there can 
be easily seen from
equation (\ref{eq:alph}) which allows us to rewrite equation 
(\ref{eq:Q_visc_irrad}) as $Q/Q_0=\alpha_\nu/\alpha_{GI}$
and this ratio is $> 1$ since for $\alpha_\nu$ to dominate 
over the gravitoturbulent torques $\alpha_\nu>\alpha_{GI}$
must be fulfilled. It also follows from equation (\ref{eq:tau_eq_irrad}) 
that the $\tau=1$ transition, if it occurs in the externally 
irradiated region of viscous disk, takes place at
\ba
\omega=\omega_1=\frac{\alpha_\nu}{\dot m}
\left(\frac{T_0}{T_f}\right)^{1-\beta}.
\ea

\section{Phase space plots.}\label{app2}

The $(\dot m,\omega)$ coordinates of various critical points in 
Figure \ref{fig:CaseA} corresponding to $T_0/T_f>1$ are
\ba
&& A=\left(\alpha_\nu\left(\frac{T_0}{T_f}\right)^{3/2},
\alpha_\nu^{-1/3}\left(\frac{T_0}{T_f}\right)^{(2-\beta)/3}\right),~~~B=
\left(\alpha_\nu\left(\frac{T_0}{T_f}\right)^{3/2},
\left(\frac{T_0}{T_f}\right)^{-(\beta+1/2)}\right),\nonumber\\
&& C=\left(\left(\frac{T_0}{T_f}\right)^{3/2},
\left(\frac{T_0}{T_f}\right)^{-(\beta+1/2)}\right),
~~~D=\left(\left(\frac{T_0}{T_f}\right)^{3/2},
\left(\frac{T_0}{T_f}\right)^{(2-\beta)/3}\right).
\label{eq:CaseA_points}
\ea
The same for $\alpha_\nu^{2/(7+4\beta)}<T_0/T_f<1$ displayed in 
Figure \ref{fig:CaseB}:
\ba
&& A=\left(\alpha_\nu\left(\frac{T_0}{T_f}\right)^{3/2},
\alpha_\nu^{-1/3}\left(\frac{T_0}{T_f}\right)^{(2-\beta)/3}\right),~~~B=
\left(\alpha_\nu\left(\frac{T_0}{T_f}\right)^{3/2},
\left(\frac{T_0}{T_f}\right)^{-(\beta+1/2)}\right),\nonumber\\
&& C=\left(1,1\right),~~~D=\left(\left(\frac{T_0}{T_f}\right)^{3/2},
\left(\frac{T_0}{T_f}\right)^{3+\beta}\right),
~~~F=\left(\left(\frac{T_0}{T_f}\right)^{5+2\beta},
\left(\frac{T_0}{T_f}\right)^{-(\beta+1/2)}\right),
\label{eq:CaseB_points}
\ea
and for $T_0/T_f<\alpha_\nu^{2/(7+4\beta)}$ shown in Figure \ref{fig:CaseC}:
\ba
&& A=\left(\alpha_\nu\left(\frac{T_0}{T_f}\right)^{3/2},
\alpha_\nu^{-1}\left(\frac{T_0}{T_f}\right)^{3+\beta}\right),~~~B=
\left(\alpha_\nu^{(10+4\beta)/(7+4\beta)},
\alpha_\nu^{-(2\beta+1)/(7+4\beta)}\right),\nonumber\\
&& C=\left(1,1\right),~~~D=\left(\left(\frac{T_0}{T_f}\right)^{3/2},
\left(\frac{T_0}{T_f}\right)^{3+\beta}\right).
\label{eq:CaseC_points}
\ea

\end{document}